\documentclass[final,twocolumn,3p,times]{elsarticle}
\usepackage{graphicx}
\usepackage{amssymb}
\usepackage{amsmath}
\usepackage{hyperref}
\usepackage{color}

\journal{Physics Letters A}
 
\begin{document}

\title{Symmetry-breaking   vortex-lattice  of a binary  superfluid in a rotating bucket}

 \author{S. K. Adhikari }
        \address{Instituto de F\'{\i}sica Te\'orica, Universidade Estadual
                     Paulista - UNESP,  01.140-070 S\~ao Paulo, S\~ao Paulo, Brazil}

 \date{\today}


\begin{abstract}

We study spontaneous-symmetry-broken phase-separated vortex lattice  in  a weakly interacting {\it uniform} 
rapidly rotating  binary Bose  superfluid % with repulsive inter- and intra-species interactions at zero temperature 
 contained in 
a quasi-two-dimensional  circular or square bucket. %  For a weak inter-species repulsion, the two component superfluids 
%remain in a mixed phase generating overlapping vortex lattices in the two components. However, 
For  the   inter-species repulsion above a critical value, the two superfluid components separate  
and form a demixed phase with practically no overlap in the vortex lattices of the two components,  which will permit an efficient experimental  observation of such vortices and study their properties. In case of a circular bucket with  equal  intra-species energies of the two components, the two components separate into two non-overlapping semicircular domains for all frequencies of rotation $\Omega$ 
generating distinct demixed vortex lattices.
% When the intra-species energies are different,
%the shape of the two components forming vortex lattice changes with $\Omega$.  
In case of a binary Bose superfluid in both circular and square buckets, (a)  the number of vortices increases linearly with $\Omega$ in agreement with a  suggestion by Feynman, and (b) the rotational energy in the  rotating frame decreases quadratically with $\Omega$ in agreement with a suggestion by Fetter. % Both suggestions of Feynman and Fetter were made for a single component superfluid.

%  We demonstrate our claim with a numerical simulation of the mean-field Gross-Pitaevskii equation for a rapidly rotating quasi-2D uniform binary Bose superfluid.

\end{abstract}

%\pacs{03.75.Hh, 03.75.Mn, 03.75.Kk}

%\begin{keyword}
%Rotating binary  Bose-Einstein condensate; Gross-Pitaevskii equation; Split-step Crank-Nicolson scheme;
%Spontaneous symmetry breaking, Vortex lattice 
%\end{keyword}

\maketitle

\section{Introduction}

The formation of a regular lattice of quantized vortices in a uniform superfluid under rotation signals  its superfluid nature. %Tisza \cite{tisza} and Landau \cite{landau} % were the pioneers in the study of superfluidity using hydrodynamical equations \cite{lf} mostly in relation to   liquid He II.
%formulated  hydrodynamic equations for a irrotational, non-viscous, frictionless superfluid to  describe many collective  properties of  He II reasonably well \cite{tf}. % From independent theoretical considerations, 
%Onsager \cite{onsager}, Feynman \cite{feynman}, and Abrikosov \cite{abrikosov} unveiled an interesting  property  that 
The integral of the velocity field $\bf v$ around a generic closed path  $\cal C$ in a rotating  superfluid  is quantized 
\cite{onsager,feynman,abri,fetter}: 
\begin{equation}\label{one} \oint_{\cal C }  {\bf v} \cdot d{\bf r}= 2 \pi \hbar l/m,
\end{equation} 
where $m$ is the mass of an atom in the superfluid, the associated angular momentum  $l$ of a  is zero or an integer (for a quantized vortex) and $m$ is the mass of an atom. % The possibility   $l\ne 0$ signals topological defect(s) inside  path $\cal C$ in the form of vortex line(s) of quantized angular momentum $l\hbar $ per atom. % As the superfluid is frictionless and irrotational, the superfluid in a rotating bucket does not rotate like a rigid body, yet develops quantized vortex lines with a density minimum. 
%As the angular frequency of rotation  $\Omega$   of the bucket is increased from zero, a single vortex line appears first as observed  and studied  by Vinen \cite{vinen,vinen2}. 
%Abrikosov \cite{abrikosov} demonstrated from
%energetic consideration that 
A rapidly rotating uniform   superfluid  prefers many vortices
of unit angular momentum  arranged in a regular lattice, often a triangular lattice,  over a
vortex of multiple angular momentum \cite{abri}. 
Quantized vortices  of unit angular momentum were first observed  in a uniform  superfluid He II in a rotating bucket  \cite{vinen,gwp}.
%Hence many vortices in a definite pattern emerge upon an increase of  \cite{sonin}  $\Omega$  as observed by 
%Gordon et al.  \cite{gwp} % photographed   actual experimental   patterns of vortices 
%in  He II in a rotating bucket. Stauffer and Fetter \cite{sf}  reported computer simulation of vortex lattice generation  in this system.   
%In addition to liquid He II \cite{heII}, vortex lattice has been observed in type II superconductors \cite{typeII}.  
%Although vortex line and patterns were first observed  in superfluid He II in a rotating bucket  \cite{gwp,vinen}, because of very strong interaction in the dense He II superfluid,  no controllable perturbative theory can be formulated for this system for a  systematic theoretical study. The number of generated vortices in these  experiments in rotating He II was small \cite{vinen}. 
  Soon after the observation of a trapped dilute weak-coupling Bose-Einstein condensate (BEC) in
alkali-metal atoms   in a laboratory \cite{becexpt}, a single vortex \cite{vors} and
 hundreds of vortices \cite{vorl}  arranged in Abrikosov triangular lattice 
in a rapidly rotating harmonically-trapped   superfluid  BEC were observed  under controlled conditions   and studied
theoretically \cite{1comp}  using the weak-coupling mean-field Gross-Pitaevskii (GP) equation \cite{gp}. 

More recently it has been possible to create a BEC in a laboratory in a one- \cite{13}
and a multi-dimensional \cite{14} optical box trap so that the condensate is subjected to
a uniform potential inside the box.
The optical trap considered in Ref. \cite{14} is a uniform three-dimensional cylindrical box trap  which combines a circular box trap in the $x - y$ plane
and a  one-dimensional confinement in the $z$ direction 
which is really the bucket of  rotating He II experiments \cite{vinen,gwp}. 
{It has been demonstrated that a rapidly rotating quasi-two-dimensional (quasi-2D) uniform scalar BEC in  circular and square buckets can form a variety of vortex lattice structures including  triangular and square lattices \cite{jpcm}.}
With this experimental possibility of realizing a uniform dilute BEC in a bucket, in this Letter,  we set to study vortex generation  in   a dilute uniform rapidly rotating  quasi-2D binary BEC in a   circular or a square bucket. { Recently, we studied triangular vortex lattice generation in a binary BEC in a harmonic trap \cite{physe}.}

The number of vortices in a rapidly rotating   large uniform  superfluid    
increases linearly with the angular frequency  of rotation $\Omega$ according to a suggestion of Feynman.
 In a harmonically trapped rotating BEC, this linear relation is grossly violated with the number of vortices rapidly increasing as $\Omega$ approaches the trapping frequency \cite{fetter}. Similarly, Fetter suggested  using general considerations that the $\Omega$-dependent  rotational energy of a single-component uniform superfluid  in a bucket
is the rotational
energy of rigid-body rotation of the superfluid with the corresponding
moment of inertia and is proportional to   $\Omega^2$.
 We find that both these laws, e.g., the linear increase of the number of vortices with $\Omega$  and quadratic decrease of the rotational energy with $\Omega$, 
are also valid in the present case of a rotating  uniform binary BEC.

%Vortex lattice formation in a trapped rotating binary BEC   with a large number of vortices has also been observed \cite{Schweikhard,Hall} and studied theoretically \cite{wang,thlatsep}.
%There has also been study of vortex-lattice formation in a BEC along the weak-coupling to unitarity crossover \cite{sci}.
 { The study of vortex lattices in a binary or a multi-component spinor BEC is interesting %because the interplay between intra-species and inter-species interactions 
because it may lead to the formation of  square    \cite{jpcm,Schweikhard,kumar}, stripe \cite{jpcm} and honeycomb 
 \cite{honey} vortex lattices, other than the standard Abrikosov triangular lattice \cite{abri},     in addition to coreless vortices \cite{coreless}, 
  vortices of fractional charge \cite{Cipriani},   and phase-separated vortex lattices in multi-component non-spinor \cite{cns}  and dipolar  \cite{kumar}   BECs. The difficulty of 
experimental study of overlapping vortices in  different components of  a multi-component  BEC is monumental and despite great interest %in the study of vortex lattices in a binary BEC 
\cite{Mueller,Kasashet},
 there are only a few   experimental   studies \cite{Schweikhard,Hall} on vortex lattices in a rotating binary BEC.  
Hence    it will be highly desirable  to have phase-separated vortex lattices in a binary BEC, where the vortices and component density  of one component do not overlap with those of the other.  In this Letter, we investigate the possibility of generating phase-separated triangular, circular  and square vortex lattice structure in a binary rotating BEC confined in a square or a circular bucket.}

In a repulsive uniform binary BEC,  phase separation takes place for  \cite{phse}
\begin{equation} \label{eq1}
\kappa \equiv \frac{g_1g_2}{g_{12}^2}<1, 
\end{equation}
where $g_1$ and $g_2$ are intra-species repulsion strengths for components 1 and 2, respectively, and $g_{12}$ inter-species repulsion strength.    
We find that  the parameter domain leading to completely phase-separated vortex lattices in a rotating quasi-2D binary BEC is governed by condition (\ref{eq1}) so that the vortices of one component do not have any overlap with the matter density of the other component. 
We  find that  if $g_1=g_2$  and  $\kappa <1$, then for a circular rotating bucket the phase separated components may have the form of completely non-overlapping semi-circles lying opposite to each other  spontaneously breaking the circular symmetry of the underlying Lagrangian. In presence of rotation, the vortices are formed 
on the phase-separated semicircular components. 
 In case of a square bucket with $g_1=g_2$ the phase-separated components have the form of adjacent rectangles spontaneously breaking the symmetry of the underlying Lagrangian.  
In case $g_1 \ne g_2$ and   $\kappa<1$,  for a  quasi-2D BEC in a circular bucket
there will be complete phase separation between components with irregular shape  and upon rotation, vortices are formed on phase separated components with spontaneous symmetry breaking. 
This will facilitate the experimental observation and consequent studies of vortex lattices in binary BEC where the vortex lattice of one component has no overlap with the other component. {It is worthwhile to point out that none of the previous studies on rotating binary BEC, identified vortex lattice in a phase-separated phase. In all previous studies overlapping configuration of the two components were considered.}

% We will use a numerical simulation  of the  coupled GP equation appropriate for the binary BEC to study vortex formation on phase-separated components in a rotating quasi-2D uniform  binary BEC in  circular and square buckets.  

%In Sec. II  the mean-field model for a   rapidly rotating binary BEC is presented.  Under a tight trap in the transverse direction a quasi-2D version of the model is also given, which we use in the present study. 
%The results of numerical calculation are shown in Sec. III.  
%Finally, in Sec. IV we present a brief summary of our findings.

\section{Mean-field model for a rapidly rotating binary BEC}

A dilute  BEC of $N$ atoms of mass $m$ each in the weak-coupling limit at zero temperature is described by the following mean-field 
 GP equation \cite{gp}
\begin{equation}\label{gpt}
{\mbox i} \hbar \frac{\partial \phi({\bf r},t)}{\partial t}= \left[-\frac{\hbar^2}{2m}\nabla^2+V({\bf r}) +\frac{4\pi \hbar^2 a}{m}
n\right] \phi({\bf r},t)
\end{equation}
where $a$ is the scattering length of atoms,   density $n=N
|\phi({\bf r},t)|^2$, and normalization $\int d{\bf r}  |\phi({\bf r},t)|^2=1$, and $N$ the number of atoms.

We will describe a binary rotating uniform  BEC  %interacting via  inter- and intra-species 
%repulsions
 in a cylindrical or square  bucket in a similar fashion using a generalization of the GP equation
(\ref{gpt}) for the binary system \cite{cns,ps}.  
The GP equation has been well tested in the generation of vortices in a harmonically trapped (non-uniform)
single component \cite{1comp}  and binary \cite{Schweikhard,Hall,binary} BEC. 
 Now it is possible to make a binary BEC of two hyper-fine states of the same atomic species, such as $^{39}$K and  $^{87}$Rb  atoms.
   In such cases   the masses of the two species are equal. Thus in this theoretical study we will take the masses of two species to be equal.  
The study of a rapidly rotating binary BEC is conveniently performed in the rotating frame, where the generated vortex  
lattice is a stationary state \cite{fetter}, which can be obtained by the imaginary-time propagation method \cite{imag}. 
Such a dynamical equation in the rotating frame can be written if we note that the Hamiltonian in the rotating frame is given by  \cite{ll1960}  $H = H_0-\Omega l_z$, where $H_0$ is that in the laboratory frame,   $l_z$ is the $z$ component of angular momentum given by $l_z= i\hbar (y\partial/\partial  x - x \partial/\partial y )$. 
With the inclusion of the extra rotational energy  $-\Omega l_z$ in the Hamiltonian,   the coupled GP
equations for the binary  BEC in the rotating frame   can be written as \cite{cns,ps}
\begin{eqnarray} \,
{\mbox i} \hbar &\frac{\partial \phi_1({\bf r},t)}{\partial t} =
{\Big [}  -\frac{\hbar^2}{2m}\nabla^2 +V({\bf r}) -\Omega l_z %\nonumber
%\nonumber
%\\  &  
+\frac{4\pi \hbar^2}{m} \Big\{{a}_1 N_1 
\nonumber
\\  & \times  \vert \phi_1({\bf r},t)\vert^2
+ {a}_{12} N_2 \vert \phi_2({\bf r},t)|^2\Big\}
{\Big ]}  \phi_1({\bf r},t),
\label{eq1x}
\\  
%\end{eqnarray}
%\begin{eqnarray}
\label{eq2}
{\mbox i} \hbar  & \frac{\partial \phi_2({\bf r},t)}{\partial t} =
{\Big [}  -\frac{\hbar^2}{2m}\nabla^2 +V({\bf r}) -\Omega l_z  
%\nonumber\\ &
+ \frac{4\pi \hbar^2}{m}\Big\{ {a}_2 N_2 
\nonumber\\ & \times \vert \phi_2({\bf r},t) \vert^2
+ {a}_{12} N_1 \vert \phi_1({\bf r},t) \vert^2\Big\}
\Big] 
 \phi_2({\bf r},t),
\end{eqnarray}
where   
 the two species of atoms   are denoted $i=1,2$, $\phi_i({\bf r},t)$ are the order parameters of the two components, 
$N_i$ is the number of atoms in species 
$i$,   ${\bf r}= \{x,y,z\},$  $a_i$ is the intra-species scattering length of species $i$,  $a_{12}$ is the inter-species scattering length. The functions $\phi_i$ are normalized as $\int d{\bf r}|\phi_i({\bf r},t)|^2 =1.$

The following   dimensionless form of Eqs.  (\ref{eq1x}) and (\ref{eq2})  can be obtained  by  the  transformation of variables: ${\bf r}' = {\bf r}/l_0,$ %l_0\equiv \sqrt{\hbar/m\omega}$, 
$t'=t\omega,  \phi_i'=   \phi_i l_0^{3/2},  \Omega'=\Omega/\omega, l_z '= l_z/\hbar$, { $\omega =\hbar/ml_0^2$}, with $l_0$  a scaling length  \cite{cns}: 
\begin{align}
{\mbox i} \frac{\partial \phi_1({\bf r},t)}{\partial t}=& \,
{\Big [}  -\frac{\nabla^2}{2 } +V({\bf r})
-\Omega l_z
+ 4\pi N_1a_1 \vert \phi_1 \vert^2 
\nonumber\\ &
+ 4\pi a_{12}N_2\vert \phi_2 \vert^2
{\Big ]}  \phi_1({\bf r},t),
\label{eq3}\\
{\mbox i} \frac{\partial \phi_2({\bf r},t)}{\partial t}=& \,{\Big [}  
- \frac{\nabla^2}{2}   +V({\bf r})-\Omega l_z
+ 4\pi N_2 a_2 \vert \phi_2 \vert^2 
\nonumber\\ &
+ 4\pi a_{12} N_1 \vert \phi_1 \vert^2 
{\Big ]}  \phi_2({\bf r},t),
\label{eq4}
\end{align} 
where for simplicity we have dropped the prime from the transformed variables.

We assume that the extension of the binary BEC mixture in the $z$ direction is limited between $z=\pm z_0$ due to strong trapping in this direction 
and the density in the $z$ direction is integrable.  The order parameter can then be written as $\phi_i({\bf r})
= \psi_i({\pmb \rho},t)\Phi_i(z)$, where the function $ \psi_i({\pmb \rho},t)$ carries the essential dynamics and $\Phi_i(z)$ is  normalizable  between $z=\pm z_0$: $\int_{-z_0}^{z_0} dz |\Phi_i(z)|^2 ={\cal N}_i$. For a strong harmonic trap in the $z$ direction $z_0 \to \infty$ and $\Phi_i(z)$ has the form of a 
Gaussian: the ground state in the harmonic trap.  In that case the 
$z$ dependence in  Eqs.  (\ref{eq3}) and (\ref{eq4}) can be integrated out \cite{luca} and we have the following quasi-2D equations
\begin{align}
{\mbox i} \frac{\partial \psi_1({\pmb \rho},t)}{\partial t}=& \,
{\Big [}  -\frac{\nabla_{\pmb \rho}^2}{2 }  + V({\pmb \rho})
   -\Omega l_z
+ g_1 \vert \psi_1 \vert^2
\nonumber\\ &
 + g_{12} \vert \psi_2 \vert^2
{\Big ]}  \psi_1({\pmb \rho},t),
\label{eq5}   \\
%\end{eqnarray}
%\begin{eqnarray}
{\mbox i} \frac{\partial \psi_2({\pmb \rho},t)}{\partial t}=& \, {\Big [}  
- \frac{\nabla_{\pmb \rho}^2}{2}+ V(\pmb \rho) -\Omega l_z 
+ g_2 \vert \psi_2 \vert^2 
\nonumber\\ &
+ g_{21} \vert \psi_1 \vert^2 
{\Big ]}  \psi_2({\pmb \rho},t),
\label{eq6}
\end{align}
where  $ {\pmb \rho}=\{ x,y\}$, $\rho^2=x^2+y^2$,
$g_i=4\pi a_i N_i {\cal M}_i/{\cal N}_i,$
%$g_2= 4\pi a_2 N_2  {\cal N}_2/{\cal M}_2  ,$
$g_{12}={4\pi } a_{12} N_2  {\cal M}_{12}/{\cal N}_1 ,$
$g_{21}={4\pi } a_{12} N_1  {\cal M}_{12}/{\cal N}_2 ,$  $\int_{-z_0}^{z_0} dz |\Phi_i(z)|^4 ={\cal M}_i$ and   $\int_{-z_0}^{z_0} dz |\Phi_1(z)|^2   |\Phi_2(z)|^2  ={\cal M}_{12}$. In this study we will take 
$g_{12}=g_{21}$ maintaining the possibility $g_1\ne g_2$.
% The quasi-2D equations  (\ref{eq5})  and (\ref{eq6}) 
%remain valid \cite{luca}
% provided that the density along the $z$ direction is integrable and due a strong trapping along $z$ direction the typical extension of the BEC in the $x-y$ plane is much larger than that in the $z$ direction.
Equations (\ref{eq5})  and (\ref{eq6}) are the mean-field GP equations for the quasi-2D rotating binary 
BEC in a bucket.   For a circular quasi-2D bucket of radius ${\cal R}$ 
\begin{eqnarray}
 V(\pmb \rho)=0, \quad   \rho < {\cal R }, \\
V(\pmb \rho)\to \infty, \quad   \rho > {\cal R },
\end{eqnarray}
and for a square bucket of lenght $d$
\begin{eqnarray}
 V(\pmb \rho)=0, \quad   |x|,|y| < {d/2 }, \\
V(\pmb \rho)\to \infty, \quad   |x|,|y| > {d/2 }.
\end{eqnarray}

The binary GP equations  (\ref{eq5}) and (\ref{eq6}) can also be obtained using a variational procedure:
\begin{equation}
 {\mbox i}\frac{\partial \psi_i({\pmb \rho},t)}{\partial t}=\frac{\delta E}{\delta \psi_i^*({\pmb \rho},t)}
\end{equation}
with the following energy functional in the rotating frame:
\begin{align}\label{en}
E(\Omega) &= \frac{1}{2}\int d {\pmb \rho} \Big[\sum_ i \frac{1}{2}   \biggr( |\nabla_{\pmb \rho} \psi_i|^2{+2V({\pmb \rho})|\psi_i|^2}
+ g_i|\psi_i|^4
\nonumber \\
&- 2\psi_i^*l_z \Omega \psi_i \biggr) 
+ g_{12}|\psi_1|^2|\psi_2|^2\Big] .
\end{align}
This energy functional is essentially the average energy of a single atom in the symmetric case: $N_1=N_2$.  
We note that the rotational  energy $ \int d {\pmb \rho}\psi_i^*l_z \Omega \psi_i $  is positive. Hence the energy $E$ in the rotating frame  will decrease with the increase of angular frequency of rotation. All other contributions to energy (\ref{en}) are positive.  The contribution of the rotational energy  will be \cite{fetter} proportional to $\Omega^2$   and the energy will decrease quadratically with $\Omega$.

\section{Numerical Results}

%The quasi-2D binary mean-field equations
% (\ref{eq5}) and (\ref{eq6}) cannot be solved analytically and different numerical methods, such as the split time-step Crank-Nicolson method \cite{imag,CPC} or the pseudo-spectral method \cite{PS}, can be employed  for their solution. 
To solve Eqs. (\ref{eq5}) and (\ref{eq6}) numerically,
we propagate   these equations  in time  by the split time-step
Crank-Nicolson discretization scheme \cite{imag,CPC}  using a space step of 0.05
and a time step of 0.0002 for imaginary-time simulation and 0.0001 for real-time simulation, respectively, to obtain the stationary state and to study the dynamics.  There are different
C and FORTRAN programs for solving the GP equation \cite{imag,CPC}  and one should use the appropriate one.
These programs have recently been adapted to simulate the vortex lattice in a rapidly rotating BEC \cite{cpckk} and we use these in this study. 
 In this Letter, without considering a specific atom, we will present the results in dimensionless units for different sets of  parameters: 
$\Omega, g_1, g_2, g_{12} (=g_{21})$. 
In the phenomenology of a specific atom, the parameters   $g_1, g_2, g_{12}$ can be varied experimentally through a variation of the underlying intra- and inter-species scattering 
lengths 
 by the Feshbach resonance technique \cite{fesh}.

\begin{figure}[!t]

\begin{center}
\includegraphics[trim = 0cm 0.0cm 0cm 0mm, clip,width=.9\linewidth]{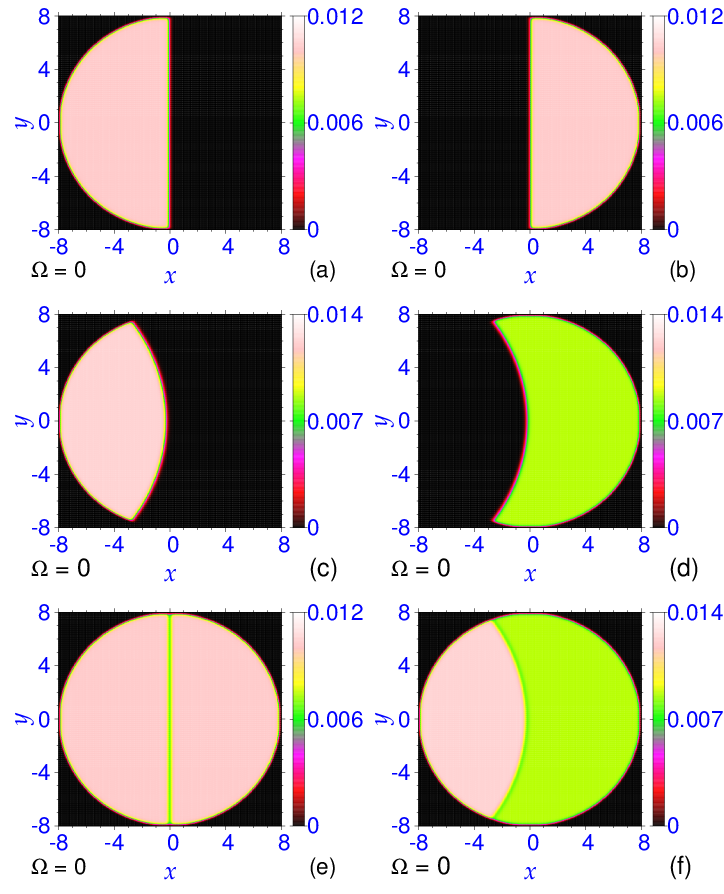} 

\caption{   Phase separation     in a non-rotating ($\Omega =0$) quasi-2D  binary BEC in a circular bucket from a contour plot of 2D densities ($|\psi_i|^2$): (a) first and (b) second components for $g_1=g_2= 5000, g_{12}=10000$, (c) first and (d) second components for 
$ g_1=5000, g_2=g_{12}=10000 $. 
The densities of the two components together for  $g_1=g_2=5000$ and and for $g_1=5000, g_2=10000$  are shown in (e) and (f),
respectively. 
 All quantities plotted in this and following figures are dimensionless.
}
\label{fig1}
\end{center}

\end{figure}

We   demonstrate a phase separation  in a fully repulsive ($g_i,g_{12}>0$) non-rotating quasi-2D   binary BEC ($\Omega=0$) in a circular bucket.
We find that a phase separation  follows  condition (\ref{eq1}).  We consider two distinct cases $g_1=g_2$ and 
$g_1 \ne g_2$. To illustrate these cases we consider (i) $g_1=g_2=5000, g_{12}=10000$ and (ii)  $g_1=5000, g_2=g_{12}=10000$ {in dimensionless units, both satisfying  condition (\ref{eq1}). All numbers reported in this Letter are in dimensionless units.}
  In the first case, the densities of the first and the second components are shown in Figs. \ref{fig1}(a) and (b), respectively.  In this case, there is a complete phase separation with the two components occupying two self-avoiding semi circles. In the second case, the densities of the first and the second components are shown in Figs. \ref{fig1}(c) and (d), respectively. In this case,
 there is also a complete phase separation, but  with the two components occupying different areas. As $g_2>g_1$, the second component occupies a larger area.     In Figs. \ref{fig1}(e) and (f) 
the densities of the two components   for cases (i) and (ii)  are  displayed together in the same plot.  In the first case in 
Fig. \ref{fig1}(e) the { interphase} between the two components is indicated by a lowering of density.
In the second case in  Fig. \ref{fig1}(f)  the two component densities are very different and the {interphase} between the two components can be easily identified. This kind of phase separation breaks the circular symmetry of the system and this symmetry breaking continues in presence of rapid rotation with vortex generation.

\begin{figure}[!t]

\begin{center}
\includegraphics[trim = 0cm 0cm 0cm 0mm, clip,width=.9\linewidth]{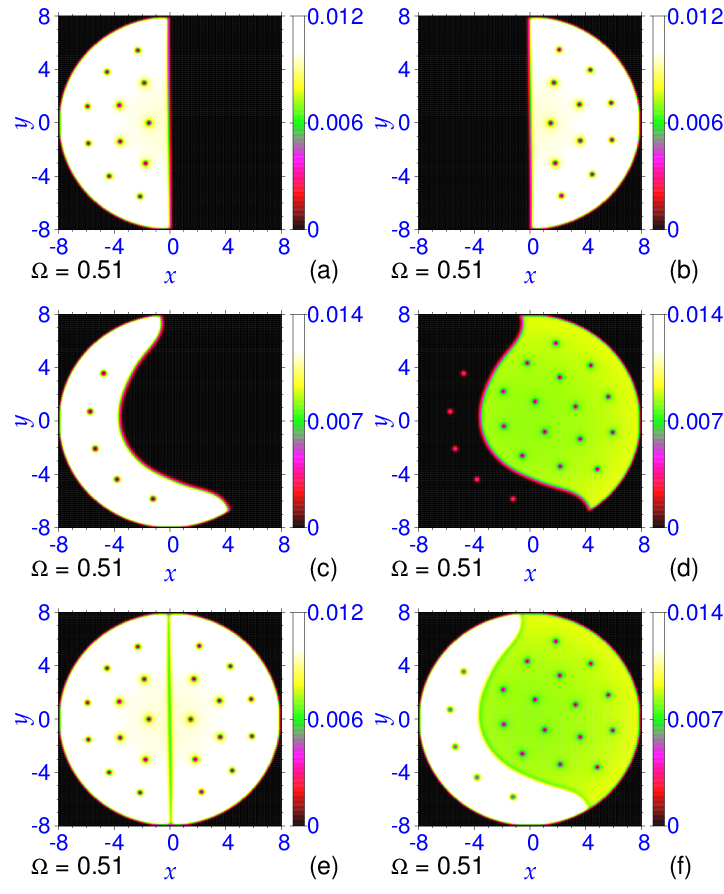} 
 
\caption{   Same as in Fig. \ref{fig1} for a rotating quasi-2D binary BEC for  angular frequency $\Omega =0.51$. }
\label{fig2}
\end{center}

\end{figure}

In Figs. \ref{fig2}(a)-(d) we plot the component densities of a rotating quasi-2D binary BEC in a circular bucket with angular frequency of rotation
 $\Omega =0.51$ with same interaction parameters $g_1,g_2, g_{12}$ as the non-rotating BECs considered in Figs.  \ref{fig1}(a)-(d). In the symmetric case illustrated in Figs. \ref{fig2}(a)-(b) ($g_1=g_2$), the components maintain the phase-separated semicircular shapes as in   Figs. \ref{fig1}(a)-(b).  Under  rotation
the same number of vortices  has appeared in  Figs. \ref{fig2}(a)-(b)  (11+11=22). However, the numbers of vortices in the two components could be different for a general arbitrary $\Omega$. 
In the asymmetric case shown in Figs. \ref{fig2}(c)-(d), the two components  remain phase separated, but the   shapes of individual components have changed when compared with those of Figs. \ref{fig1}(c)-(d), where different numbers of vortices (5+15=20) have appeared in the two components.   
{ The change of shape of the two component densities in Figs. \ref{fig2}(c)-(d) upon rotation 
 is due to the nonuniform centrifugal force acting on the two components. The average distance of the center of mass of the  first component from the center, viz. Fig.  \ref{fig1}(c), is larger than that of the second conponent from the center, viz. Fig. \ref{fig1}(d) resulting in a larger centrifugal force on the former upon rotation. Hence the first component is sqeezed more than the second component upon rotation due to a larger centrifugal acceleration,  resulting in a inner concavity  of the first component and a convexity of the second. 
 In Figs. \ref{fig1}(a)-(b) the sizes of the two components are same and subject to the same centrifugal force. Hence their shapes do not change upon rotation.  }
In Figs. \ref{fig2}(e)-(f), as in Figs. \ref{fig1}(e)-(f),  we have illustrated both components  of the rotating binary mixture in the same plot. Again, as in the non-rotating case, it is easy to identify the two self avoiding components with multiple vortices in these combined plots. Hence for a compact presentation, we will exhibit the two component densities of the rotating binary mixture  in the same plot in the following.
{ Often due to atomic repulsion, quite naturally,  the vortex core of one component is filled with atoms of the other component, as can be seen prominently in Fig. \ref{fig2}(d). }

\begin{figure}[!t]

\begin{center}
\includegraphics[trim = 0cm 0cm 0cm 0mm, clip,width=.9\linewidth]{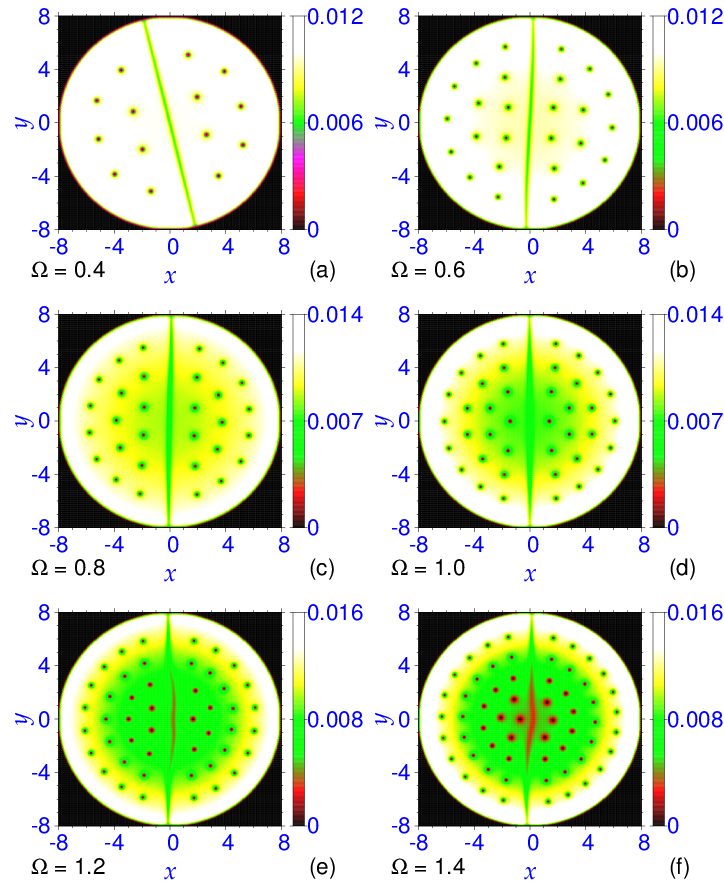} 
 
\caption{   Phase-separated vortex lattices in a rapidly rotating quasi-2D  binary BEC in a circular bucket  with $g_1=g_2=5000, 
g_{12}=10000$ for  $\Omega =$ (a) 0.4, (b) 0.6, (c) 0.8, (d) 1.0, (e) 1.2, and (f) 1.4
 from a contour plot of 2D densities ($|\psi_i|^2$).  }
\label{fig3}
\end{center}

\end{figure}

Now we study how the number of vortices  evolve with the increase of angular frequency of rotation $\Omega$ in the symmetric case $g_1=g_2 =5000$ with $g_{12}=  10000$.  We consider a large $g_{12}$, as for a large $g_{12}$, the phase separation is robust,  leading to a ground state with phase-separated vortex lattice.  For 
$g_{12}\gtrapprox g_1=g_2$, there is a phase separation, and  the ground state of a harmonically trapped binary BEC  has a phase-separated sheet structure \cite{honey,Kasashet}. 
 In Figs. \ref{fig3}(a)-(f) we display the fully phase-separated vortex-lattice states with increasing angular frequency of rotation: $\Omega= (a)  0.4, (b) 0.6, (c)  0.8, (d)  1.0, (e) 1.2$ and  (f) 1.4. The corresponding total number
of vortices in the two components are (a) 15 (=7+7), (b) 26 (=13+13), (c) 30 (=15+15), (d) 40 (=20+20), (e) 47 (=24+23), and  (f) 60 (=30+30); the numbers in the parenthesis are the numbers of vortices  in the first and the second components, respectively.   The separation between the two components can be clearly seen in this figure in a domain of low density between the two components. The maximum density in the plots in Figs. \ref{fig3}
increases with $\Omega$. We could also find for $\Omega=1.2$ two states with (23+23) and (24+24) vortices. But these states have energies greater than the (24+23) vortex state shown in Fig. \ref{fig3}(e) and hence are excited states. The  energy of a vortex-lattice state depends on the number of vortices as well as their arrangement in space. 
The vortices in Figs. \ref{fig3} are arranged in successive orbits around the center, the outer orbits are mostly circular. { In a retating harmonically-trapped binary BEC the vortices are arranged in triangular lattice \cite{physe}.}
 Although, the nonlinearities $g_i$  of the two components are the same, the number of vortices in the two components and 
the arrangement of vortices in the two components could be different.   In plots of Figs. \ref{fig3}(a)-(d) and (f)
the two components have the same number of vortices; in (e) the numbers of vortices in the two components are different.   The arrangements of the vortices in the two components 
are also the same in (a)-(d), whereas the same is different in (f).  Numerically we also identified for $\Omega= 1.4$ another degenerate ground state with the same number of vortices (30) in the two components but with the arrangement of vortices exchanged in the two components in relation to plot (f).   

\begin{figure}[!t]

\begin{center}
\includegraphics[trim = 0cm 0cm 0cm 0mm, clip,width=.9\linewidth]{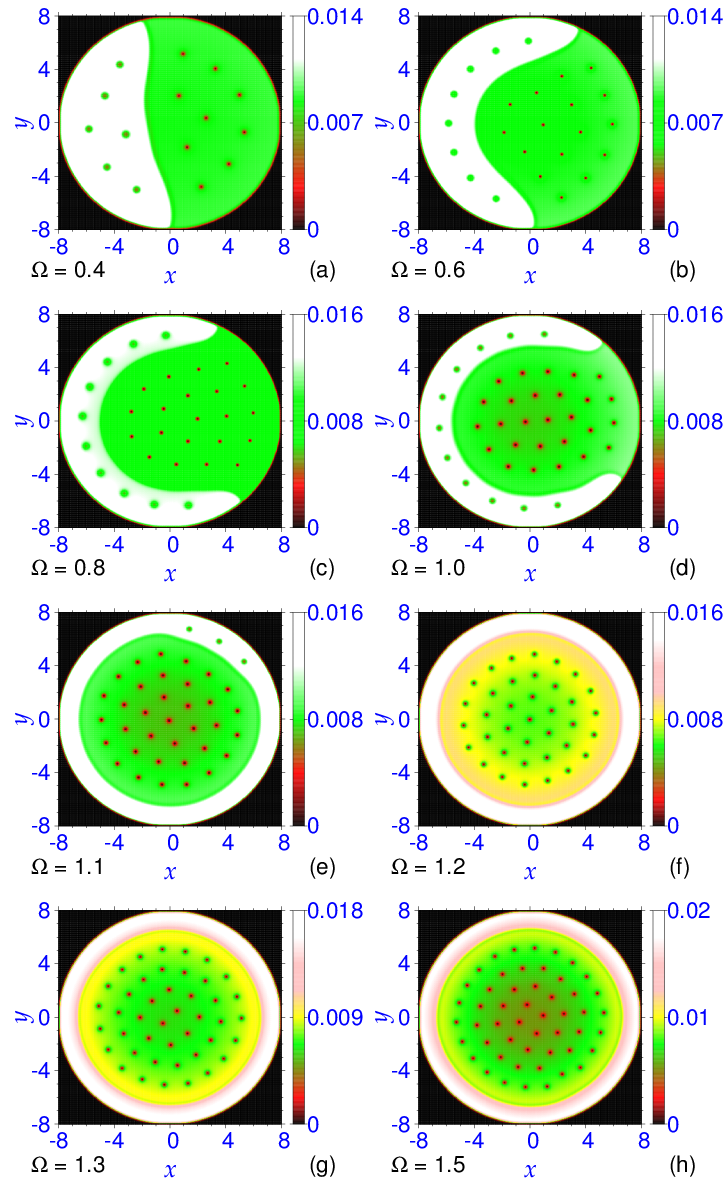}

\caption{   Phase-separated vortex lattices in a rapidly rotating  quasi-2D binary BEC in a circular bucket  with $g_1=5000, 
g_2=g_{12}=10000$ for   $\Omega = $  (a) 0.4, (b) 0.6, (c) 0.8, (d) 1.0, (e) 1.1, (f) 1.2, (g) 1.3  and (h) 1.5,   
 from a contour plot of 2D densities ($|\psi_i|^2$).  The light (pink,white) colored domain is the first component, whereas the dark (green,yellow) colored region is the second component. For a comprehensive presentation of the plots with changing density, the color scheme for plots (a)-(e) is different from plots (f)-(h).  }
\label{fig4}
\end{center}

\end{figure}

\begin{figure}[!t]

\begin{center}
\includegraphics[trim = 0cm 0cm 0cm 0mm, clip,width=.9\linewidth]{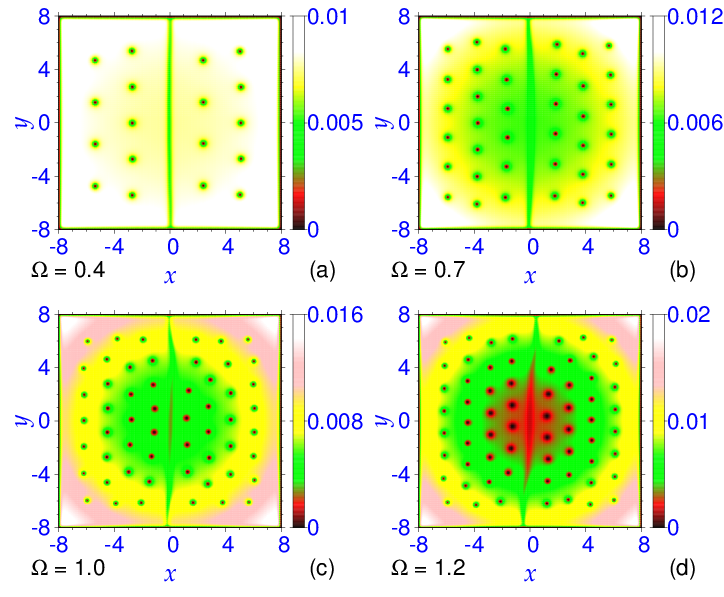} 

\caption{   Phase-separated vortex lattices in a rapidly rotating  quasi-2D binary BEC in a square bucket  with $g_1=g_2=5000, 
g_{12}=10000$ for   $\Omega = $  (a) 0.4, (b) 0.7, (c) 1.0, and (d) 1.2,  
 from a contour plot of 2D densities ($|\psi_i|^2$).   For a comprehensive presentation of the plots with changing density, the color scheme for plots (a)-(b) is different from plots (c)-(d).}
\label{fig5}
\end{center}

\end{figure}

Next we consider the evolution of  the  number of vortices  with the increase of angular frequency of rotation $\Omega$ in the asymmetric case $(g_1\ne g_2)$
with $g_1=5000$ and  $g_2=g_{12}=  10000$.    With these values of $g_i$ and $g_{12}$ the phase separation is very stable. 
 In Figs. \ref{fig4}(a)-(h) we display the fully phase-separated vortex-lattice states with increasing angular frequency of rotation: $\Omega=$ (a)  0.4, (b) 0.6, (c)  0.8, (d)  1.0, (e) 1.1, (f) 1.2, (g) 1.3  and (h) 1.5. The corresponding total number
of vortices in the two components are (a) 15 (=6+9), (b) 24 (=8+16), (c) 31 (=10+21), 
(d) 38 (=11+27), (e) 38 (=3+35), (f) 37 (=0+37), 
 (g) 43  (=0+43) and (h) 51 (0+51); again the numbers in the parenthesis are the numbers in the first and the second components,  respectively. The number of vortices in the first component increases with $\Omega$ up to about $\Omega \approx 1$. The first component surrounds the second component
more and more  as  $\Omega$ increases to about $\Omega \approx 1$. Beyond $\Omega =1$ the first component surrounds  the second component completely in the form of {a ring} around the second component occupying an inner circle and the number of vortices in the first component becomes 0 in the ground state.
 Between   $\Omega\approx 1.0$ and $\Omega\approx 1.2$ the total number of vortices  remain  fixed approximately at 37, beyond which the number of vortices in the second  component increases with $\Omega$ whereas that in the first component remains 0. The circular symmetry is broken for small $\Omega$ and is  restored for $\Omega \gtrapprox 1.2$.  In this study we displayed the lowest-energy ground states of the system. Imaginary-time simulation for $\Omega >1.2$   may lead to states with a few vortices in the first component. These states have higher energy and are discarded. 
The vortices are arranged in successive orbits in plots of Figs. \ref{fig4}(e)-(h).  The outer orbits have the shape of circles whereas the inner orbits have the shape of polygons. 
In case of a harmonically trapped BEC, similar orbits have the shape of hexagons \cite{cpckk} accommodating 1. 6, 12, 18, 24 ... vortices in successive orbits.
 In (h) the central spot at the origin does not have a vortex, while in (e) and (f) there is a vortex at the center. In (f)  the numbers of vortices in successive orbits are  1, 6, 12, 18,  as in closed hexagonal orbits  \cite{cpckk}.

\begin{figure}[!t]

\begin{center}
\includegraphics[trim = 0cm 0cm 0cm 0mm, clip,width=.9\linewidth]{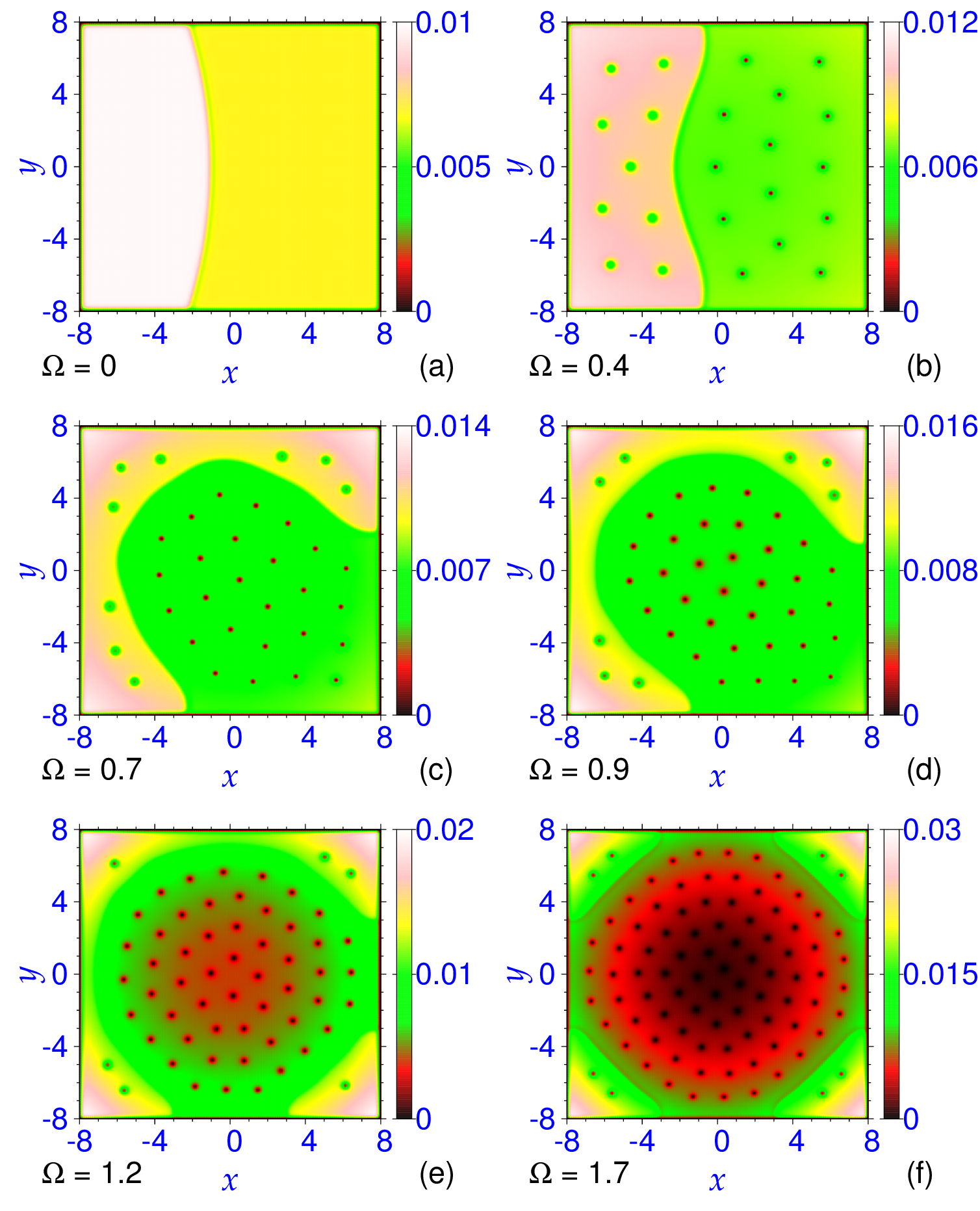}

\caption{   Phase-separated vortex lattices in a rapidly rotating  quasi-2D binary BEC in a square bucket  with $g_1=5000, 
g_2=g_{12}=10000$ for   $\Omega = $  (a) 0.4, (b) 0.7, (c) 1.0, and (d) 1.2,  
 from a contour plot of 2D densities ($|\psi_i|^2$).   For a comprehensive presentation of the plots with changing density, the color scheme for plots (a)-(b) is different from plots (c)-(d).}
\label{Fig5}
\end{center}

\end{figure}

The phase separation with symmetry breaking of a rapidly rotating quasi-2D binary BEC in a square  bucket is studied next for
 $g_1=g_2=5000$ and $g_{12}=10000$.  In Figs. \ref{fig5}(a)-(d) we display the phase-separated vortex-lattice 
states for angular frequencies $\Omega=$ (a)  0.4,  (b) 0.7, (c) 1.0, and (d)  1.2.  The corresponding total number of vortices in the two components are (a) 18 (=9+9), (b) 37 (=19+18), (c) 55 (=27+28), (d) 64 (=32+32), respectively. The number of vortices in the two components are the same in (a),  and (d) and different in (b) and (c). Except near the boundary the vortices are arranged in  triangular lattice in all plots.
In this case the number of vortices also increases with $\Omega$.

{ Next we investigate  
the phase separation  of a rapidly rotating quasi-2D binary BEC in an asymmetric  square  bucket for
 $g_1=5000$ and $g_2=g_{12}=10000$.  In Figs. \ref{Fig5}(a)-(f) we display the phase-separated vortex-lattice 
states for angular frequencies $\Omega=$ (a)  0,  (b) 0.4, (c) 0.7,  (d)  0.9, (e) 1.2, and (f) 1.7.  The corresponding total number of vortices in the two components are (a) 0, (b)  23 (=9+14), (c) 35 (=9+26), (d) 43 (=8+35), (e) 59 (=6+53), (f) 92 (=8+84), respectively. The total number of vortices increase with $\Omega$. In the central region the vortices are arranged in approximate triangular lattice.}

The number of vortices in a rotating BEC in a bucket can be obtained from a theoretical estimate of Feynman \cite{feynman}.
For a large number of vortices, the areal density of vortices (number of vortices per unit area) tends to 
\begin{equation}
{\cal N}= \frac{\Omega}{\pi} 
\end{equation}
in units $\hbar=m=1$. Hence the number of vortices in a circular area of radius $\cal R$ is
\begin{equation}\label{c}
{\bf N}_{\mbox{circle}}= {\cal R}^2 \Omega, 
\end{equation}
and that in a square of side $d$  is
\begin{equation}\label{d}
{\bf N}_{\mbox{square}}= \frac{d^2 \Omega}{\pi}.  
\end{equation}
The Feynman estimates for generated vortices (\ref{c}) and (\ref{d})  are proportional to $\Omega$  and give an idea about the number of vortices in an actual numerical calculation.

The $\Omega$-dependent part of energy per atom can be obtained from a theoretical estimate of Fetter \cite{fetter}:
\begin{equation}
E(\Omega)-E(0) = - \frac{1}{2}I \Omega^2,
\end{equation}
where $I$ is the moment of inertia of rigid-body rotation of an atom of the superfluid. For a quasi-2D BEC atom of mass $m=1$, the moment of inertia of a circle of radius ${\cal R}$ is ${\cal R}^2 /2$ and that of a {{\it uniform}} square of side $d$ is $d^2 /6$.  Hence the Fetter estimates of $\Omega$-dependent energies per atom of a BEC in a circular  and square buckets are, respectively
\begin{eqnarray}\label{ec}
E(\Omega)-E(0) =-\frac{{\cal R}^2 \Omega^2}{4}  \\
E(\Omega)-E(0) =- \frac{d^2\Omega^2}{12}.  \label{es}
\end{eqnarray}   
{ In case of the square, the estimate is applicable in the symmetric case only ($g_1=g_2$), where the density is uniform, and not in the asymmetric case ($g_1\ne g_2$) of nonuniform density. }

\begin{figure}[!t]

\begin{center}
\includegraphics[trim = 0cm 0.cm 0cm 0mm, clip,width=.8\linewidth]{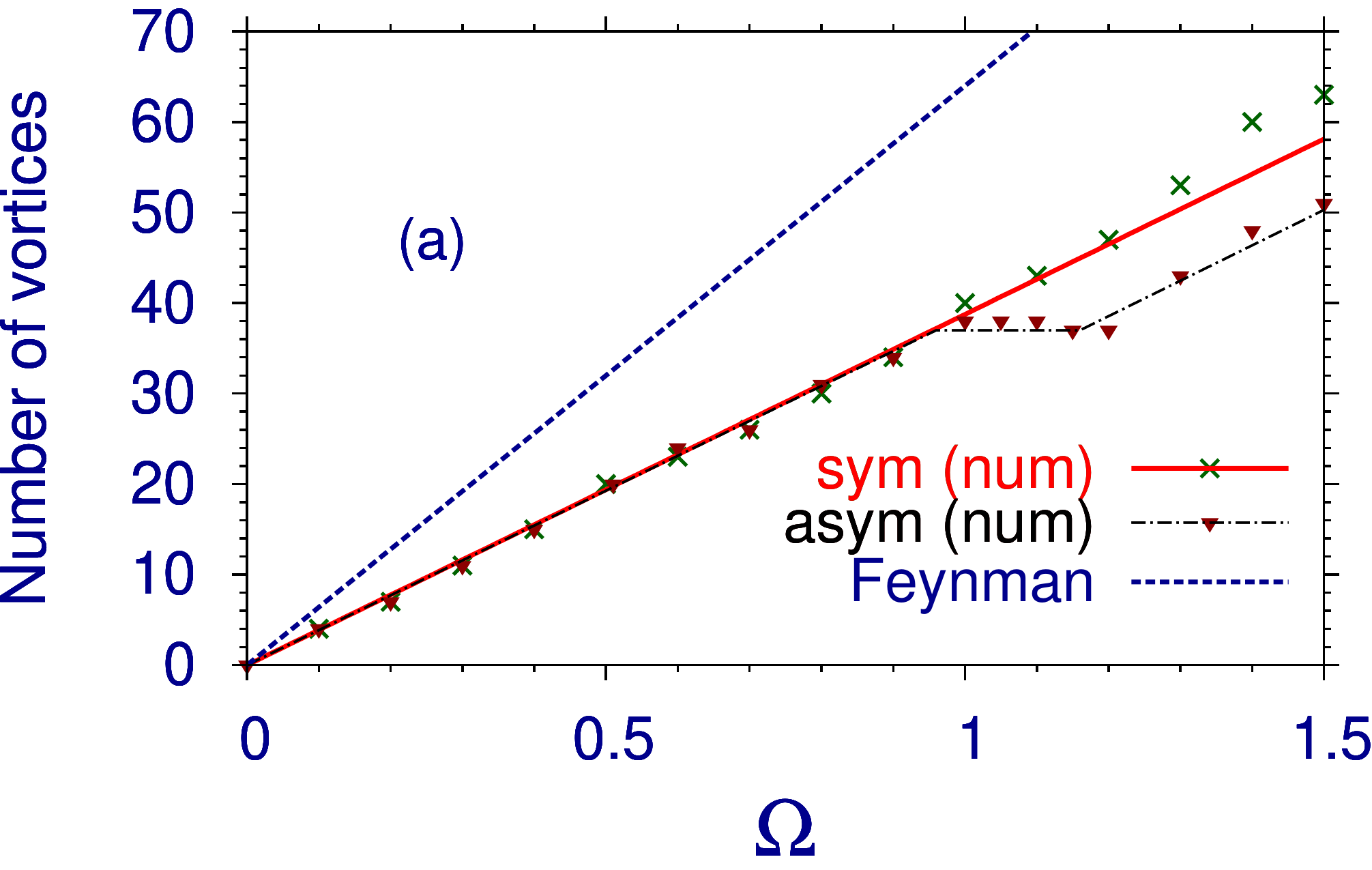}
\includegraphics[trim = 0cm 0.cm 0cm 0mm, clip,width=.8\linewidth]{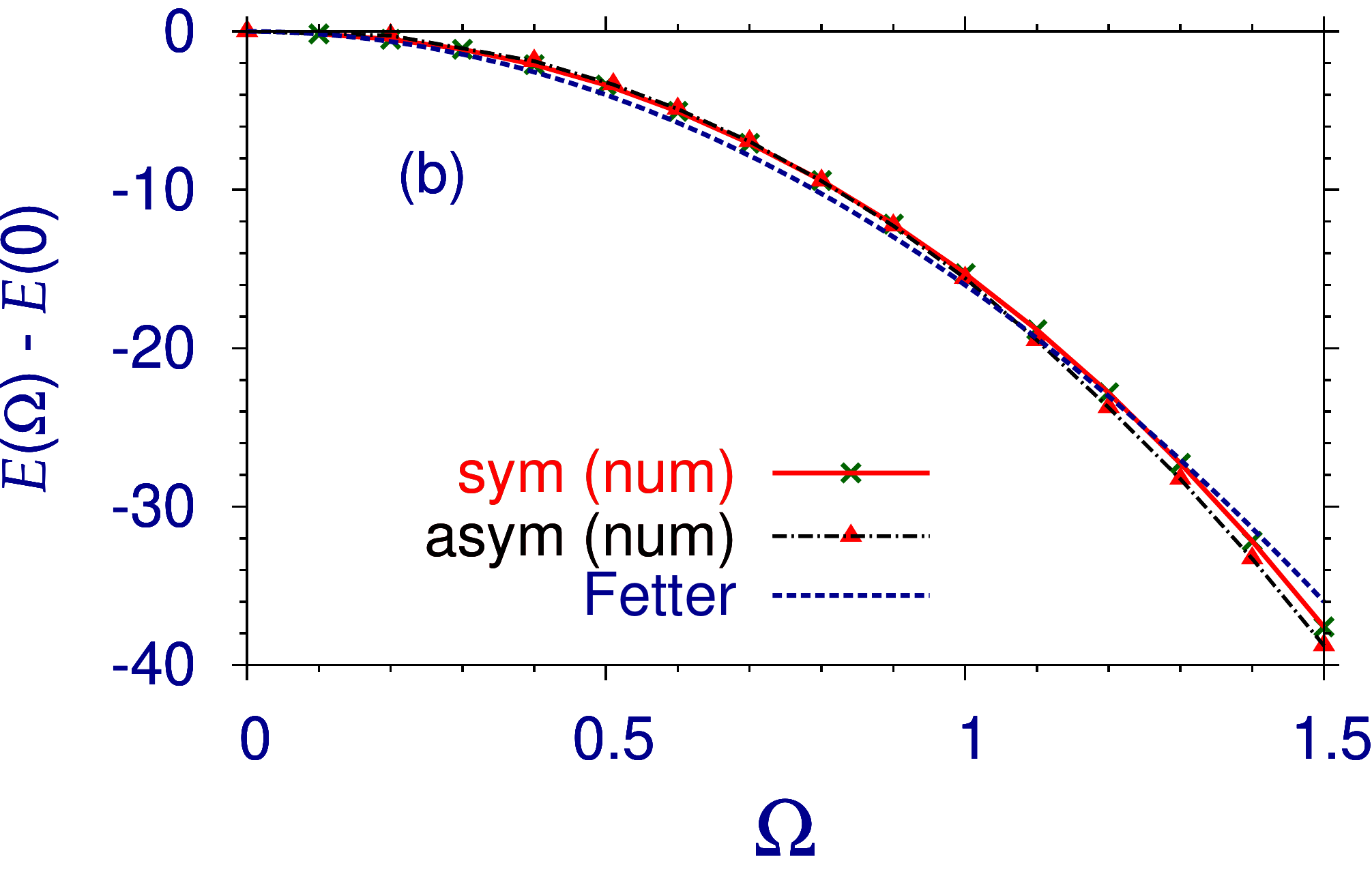} 

\caption{  (a) Number of vortices and (b)  energy $E(\Omega)-E(0)$  in the rotating frame for a
rapidly rotating quasi-2D binary BEC  in a circular bucket   versus angular frequency of rotation $\Omega$ for (i) the symmetric  (sym) case with 
$g_1=g_2=5000,g_{12}=10000$ and (ii) the asymmetric (asy) case with  $g_1=5000, g_2=g_{12}=10000$; the numerical results are labeled sym (num) and asym (num), respectively. The theoretical estimates of Feynman
(\ref{c}) and of Fetter (\ref{ec}) for the number and energy, respectively,  are also shown. 
}
\label{fig6}
\end{center}

\end{figure}

  In Figs. \ref{fig6}(a)-(b) we plot the numerically obtained  number of vortices and the $\Omega$-dependent energy in the rotating frame versus $\Omega$, respectively, for a rotating quasi-2D binary BEC in a circular bucket. We also plot the theoretical estimates of Feynman \cite{feynman}  (\ref{c}) for the number of vortices and of Fetter \cite{fetter} (\ref{ec})  for the $\Omega$-dependent part of energy in the rotating frame.   As expected, the number of vortices increases with $\Omega $ and energy decreases as $\Omega$ is increased. The energy decreases as the contribution of the rotational energy  $-\Omega l_z$ in the expression for energy (\ref{en}) is negative. 
{ In the asymmetric case around $\Omega \sim 1.0 $ to 1.2,  the first component forms a thin ring around the second component. Because of the small width of the first component it cannot accommodate any vortex. 
Thus 
 the number  of vortices in the first component reduces to zero and this volume is excluded from vortex formation, while the number of vortices in the second component continue to increase, so that the total number vortices remain constant, viz. Fig. \ref{fig6}(a). As $\Omega $ is further increased beyond $\Omega= 1.2$ the number of vortices in the second component increases linearly, whereas that in the first component remains zero in the ground state. This phenomenon does not take place in the symmetric case. 
As a consequence, the number of vortices is reduced in the asymmetric case compared  to the symmetric case, viz.  Fig. \ref{fig6}(a).   }

\begin{figure}[!t]

\begin{center}
\includegraphics[trim = 0cm 0.cm 0cm 0mm, clip,width=.8\linewidth]{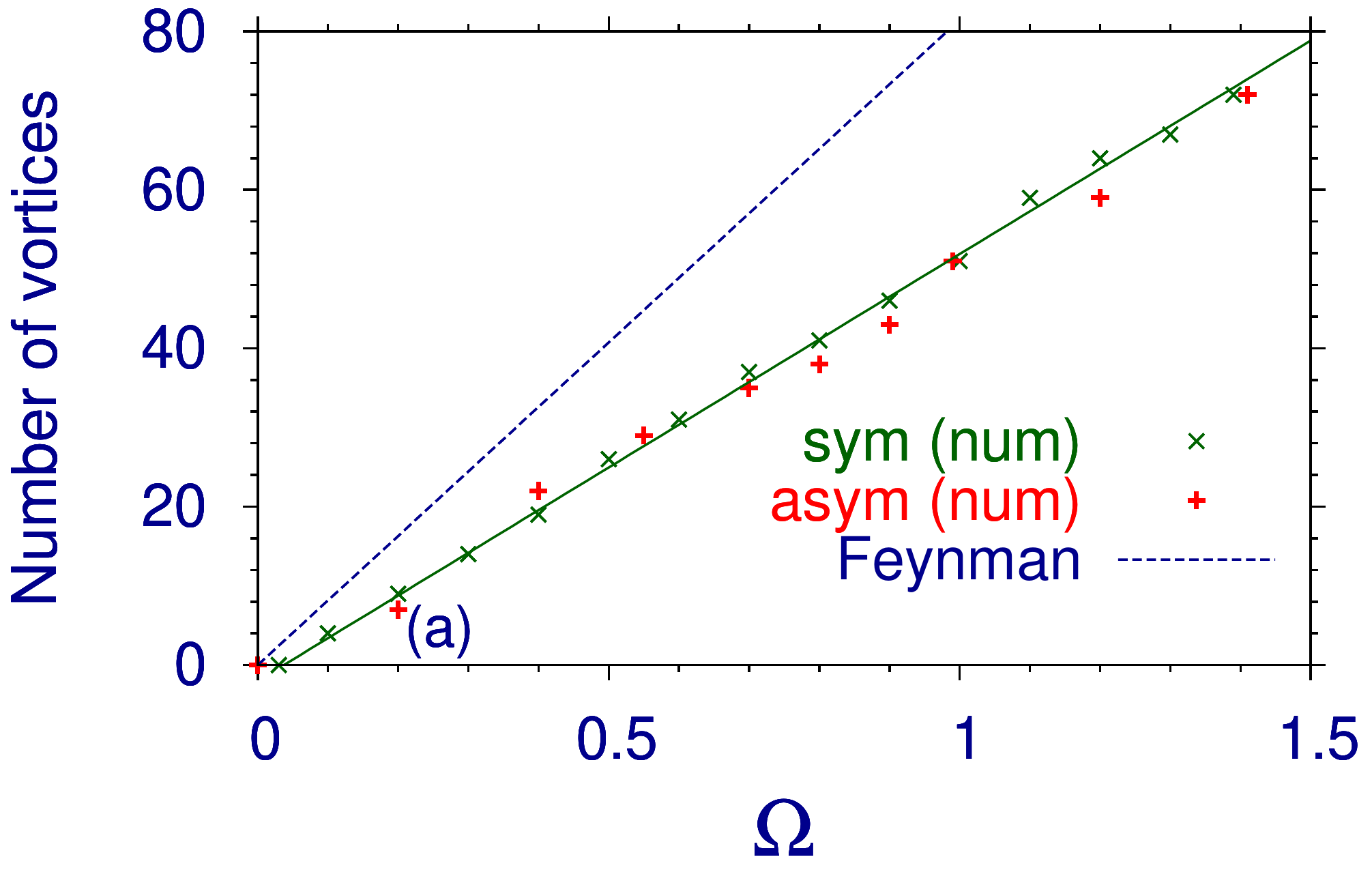}
 \includegraphics[trim = 0cm 0.cm 0cm 0mm, clip,width=.8\linewidth]{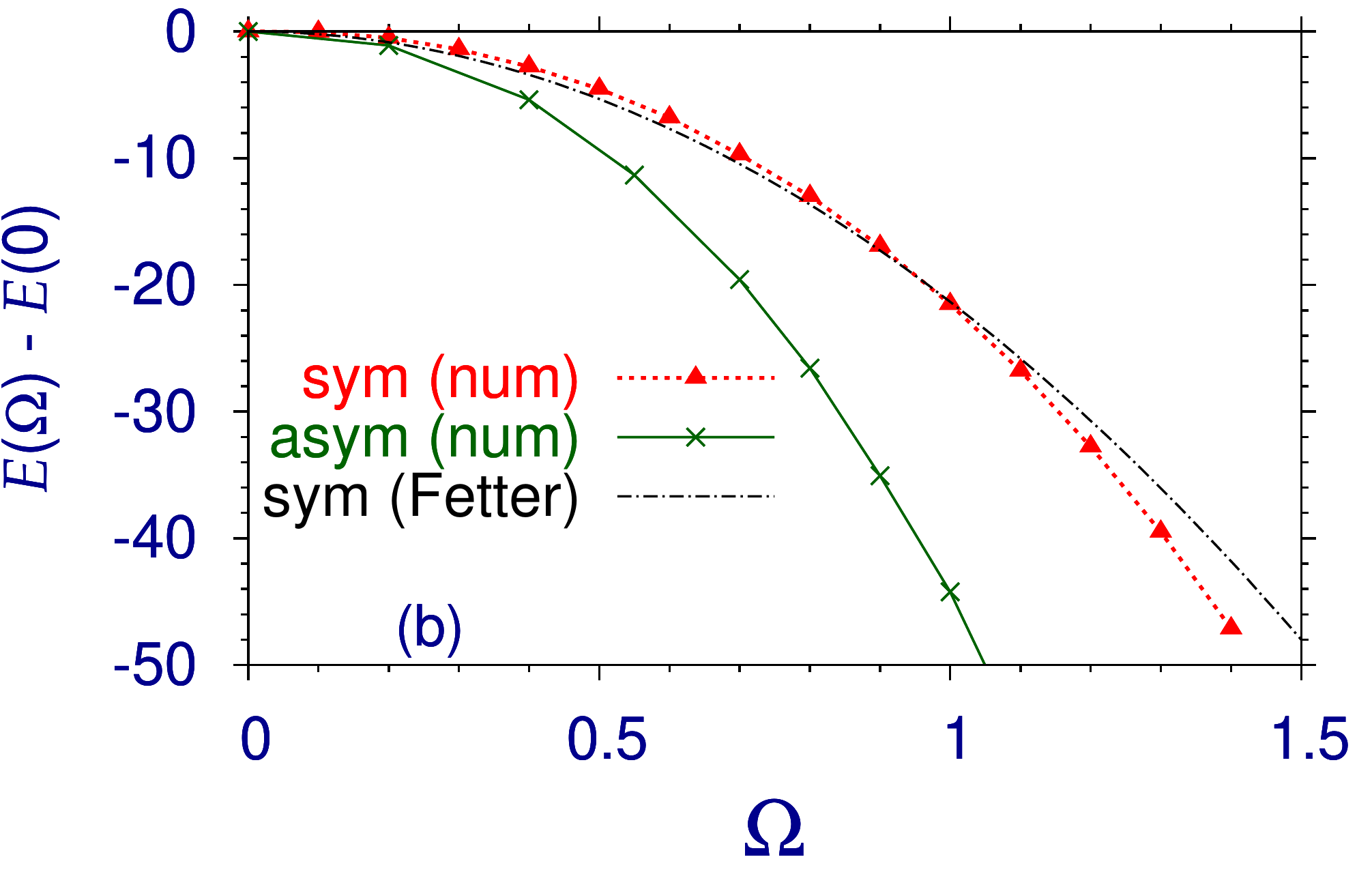}

\caption{ Numerical results for the  (a) number of vortices and (b)  energy $E(\Omega)-E(0)$  in the rotating frame for a
rapidly rotating quasi-2D binary BEC  in a square  bucket   versus angular frequency of rotation $\Omega$ for (i) the symmetric  (sym) case with 
$g_1=g_2=5000,g_{12}=10000${, and (ii) the asymmetric (asym) case with $g_1=5000,g_2=g_{12}=10000$.}  The theoretical estimates for the  number of vortices (\ref{d})  by Feynman and for rotational energy (\ref{es})  by Fetter {for the symmetric case } are also shown. { In (a) the full line through the numerical points is shown to guide the eye.}
}
\label{fig7}
\end{center}

\end{figure}

 In Figs. \ref{fig7}(a)-(b) we display the numerically obtained  number of vortices   and the $\Omega$-dependent energy in the rotating frame   versus $\Omega$, respectively, for a rotating quasi-2D binary BEC in a square  bucket and compare these with the estimate of Feynman for the number of vortices (\ref{d})  and of Fetter for energy (\ref{es}), respectively. Actually, the estimate of Feynman for the number of vortices is valid in a large system. In a small system of area $16\times 16 $ with boundary as studied in this Letter, a certain area near boundary   is excluded from the formation of vortices and the actual number of vortices should be less than the Feynman estimate.

 \begin{figure}[!t]
  
\begin{center}
\includegraphics[trim = 0cm 0.cm 0cm 0mm, clip,width=.9\linewidth]{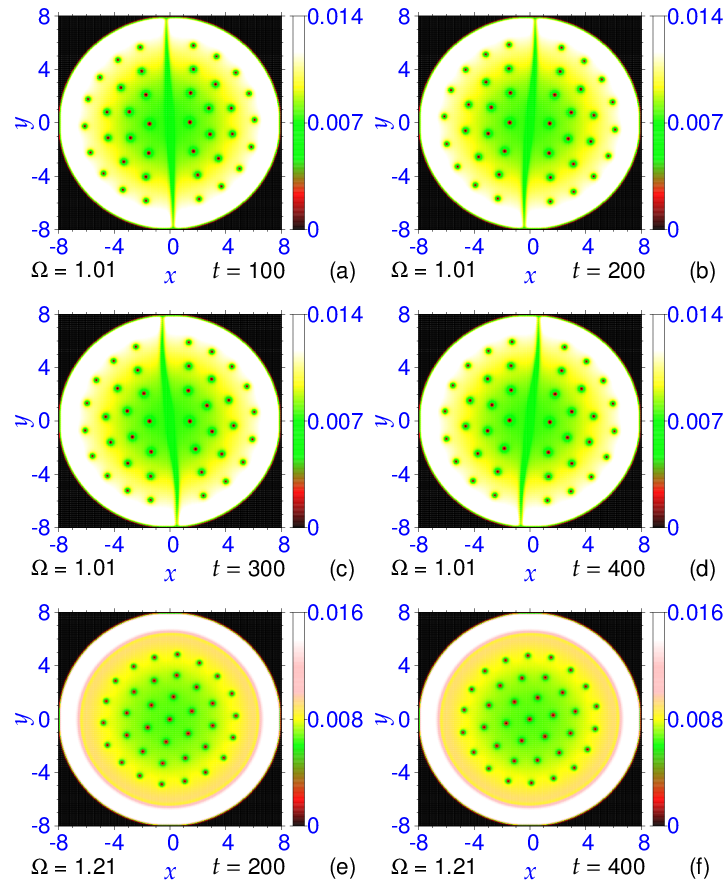}

\caption{Dynamical evolution of vortex lattice of the rotating BEC in a circular 
bucket, displayed in Fig.  \ref{fig3}(d), during real-time propagation for 400 units of time
using the corresponding imaginary-time wave function as input, at times (a)
$t = 100$, (b) $t = 200$, (c) $ t = 300,$ and (d) $t = 400$. During real-time propagation
the angular frequency of rotation $\Omega$  was changed at $t = 0$ from the imaginary-time
value of $\Omega = 1.0$ to   1.01. The same for the vortex-lattice state of  Fig.  \ref{fig4}(f)
at times (e) $t = 200$ and (f) $t = 400$.  During real-time propagation $\Omega$  was changed at $t = 0$
from 1.2 to 1.21.  
}
\label{fig8}
\end{center}

\end{figure}

The dynamical stability of the vortex-lattice states of  a quasi-2D  rotating binary BEC in a circular bucket  is tested next. First we consider the symmetric  vortex-lattice state of Fig. \ref{fig3}(d) with $g_1=g_2=5000$ and $g_{12}=10000$.
For this purpose we subject this vortex-lattice state  obtained by imaginary-time simulation to real-time simulation for a long period of time after changing the rotational frequency $\Omega$ from 1.0 to 1.01 at time $t=0$. The
vortex lattice will be destroyed after some time, if the underlying BEC wave function were dynamically
unstable.  The snapshots of subsequent evolution
of the vortex-lattice state  is displayed in Fig. \ref{fig8}  at  (a)  $t = 100$, (b) $t = 200$, (c) $t = 300$,
and (d) $t = 400.$  
 The robust nature of the snapshots of the vortex-lattice profiles  during
real-time evolution upon a small perturbation, as exhibited in Fig. \ref{fig8}, demonstrates
the dynamical stability of the vortex lattice in the quasi-2D rotating binary BEC of Fig. \ref{fig3}(d). Next we consider    the dynamical stability of the vortex lattice profile of Fig. \ref{fig4}(f) in the asymmetric case   with $g_1=5000,$ $g_2=g_{12}=10000$ and $\Omega=1.2$.   We subject this vortex-lattice state to real-time simulation for a long period of time after changing the rotational frequency $\Omega$ from 1.2 to 1.21 at time $t=0$. The snapshots of subsequent evolution
of this vortex-lattice state  is displayed in Fig. \ref{fig8}  at  (e)  $t = 200$, and  (f) $t = 400$.  The robust nature of the vortex lattice states in (e) and (f) demonstrates the dynamical stability of the vortex-lattice state of Fig. \ref{fig4}(f).

\begin{figure}[!t]
\begin{center}
\includegraphics[trim = 0cm 0.cm 0cm 0mm, clip,width=.9\linewidth]{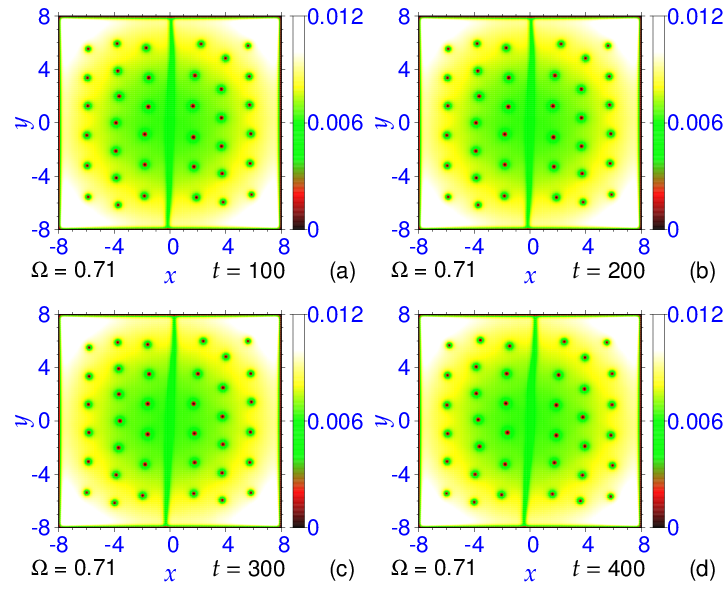} 

\caption{Dynamical evolution of vortex lattice of the rotating BEC in a square
bucket, displayed in Fig.  \ref{fig5}(b), during real-time propagation for 400 units of time
using the corresponding imaginary-time wave function as input, at times (a)
$t = 100$, (b) $t = 200$, (c) $ t = 300,$ and (d) $t = 400$. During real-time propagation
the angular frequency of rotation $\Omega$  was changed at $t = 0$ from the imaginary-time
value of $\Omega = 0.7$ to   0.71. 
}
\label{fig9}
\end{center}

\end{figure}

The vortex-lattice states of a rotating quasi-2D  binary BEC in a square bucket  are also found to be dynamically stable. For our demonstration we consider the state of  Fig. \ref{fig5}(b) with $g_1=g_2=5000, g_{12}=10000$ and $\Omega = 0.7$ as obtained by imaginary-time evolution and subject it to long-time real-time evolution after changing $\Omega$ from 0.7 to 0.71. 
In Fig \ref{fig9} we illustrate the real-time profiles of the vortex-lattice states at times $t=$ (a) 100, (b) 200, (c) 300, and (d) 400. The robust nature of these states demonstrate the dynamical stability.

\section{Summary and Discussion} 
 
We  studied the generation of spontaneous symmetry-breaking completely   phase-separated   vortex lattices in a uniform repulsive quasi-2D  binary BEC in a circular and a square box.   In the examples studied in this Letter, there is no overlap between the component densities of the BEC, so that vortex lattices of the two components are formed in different regions of space, which is of great phenomenological interest.  This will facilitate the experimental study of the vortex lattices of the two components.   
Such vortex-lattice structure  is generated when  $\kappa$ of   Eq. (\ref{eq1}) is much smaller  that unity, e.g., $\kappa \lessapprox 0.75$ or so.  For larger values of the ratio  $\kappa$ ($ 0.8 \lessapprox \kappa <1$), although there could be a phase separation for a  non-rotating binary BEC,  overlapping structure is found under rapid rotation. In this study we fixed this ratio in the symmetric case $g_1=g_2=5000$ as: $\kappa = 0.25$ and study the generation vortex lattice in the case of circular and square buckets.  For the asymmetric case $g_1=5000, g_2=10000$ we consider $\kappa =0.5$.  In all cases it was possible to have 
symmetry-breaking  phase-separated vortex lattice under rapid rotation.   In the  symmetric case, the vortex lattices of the two components lie on fully separated semicircles for a binary BEC in a circular bucket breaking the circular symmetry spontaneously.  In case of a square bucket, the phase-separated vortex lattice also breaks the symmetry of the underlying Lagrangian.
 In the asymmetric case,  although  the phase separation is complete in a circular bucket, the shapes of the two components are different. In this case, the phase-separated state breaks the symmetry for small $\Omega$, and symmetry is restored for large $\Omega$.
  We demonstrated dynamical stability of all types of vortex-lattice states  by  real-time simulation over a long period of time upon a small change in the angular frequency of rotation, viz. Figs. \ref{fig8} and \ref{fig9}.

{The GP equation is valid in the weak-coupling regime for $\kappa \equiv n^{1/3} a \ll 1$. In a recent study we demonstrated that the GP equation is valid for $\kappa <0.2 $ \cite{gautam}. It is partinent to ask if the present quasi-2D nonlinearities $g$ of $5000\sim 10000$  correspond to realistic 
values of experimental parameters, such as  atom number $N$, scattering length $a$ etc, and if the GP equation remains valid for these parameters. The answers are affirmative for a binary mixture of two hyperfine states of $N= 40000 $ $^{87}$Rb atoms of inter- and intra-species scattering lengths of about 5 nm  \cite{scatl}.  For an approximate estimate, we consider a quasi-2D harmonic oscillator trap  with trap anisotropy $\omega_x:\omega_y:\omega_z :: 1:1:100$. 
 For a scaling length $l_0= 1$ $\mu$m, corresponding to a trapping frequency of 
$\omega \approx 2\pi \times 117 $ Hz,
the three-dimensional nonlinearity $4\pi N a \approx  2500$  corresponds to a quasi-2D nonlinearity $g_{12}=2500/\sqrt{2\pi/100}\approx 10000$ \cite{luca}. We numerically solved the GP equation in three-dimensions in this case for a nonlinearity of 2500  and found a central density   $n\approx 10^3 $ $\mu$m$^{-3}$  =  10$^{15}$ cm$^3$, so that $\kappa =  n^{1/3} a  \approx 0.05 < 0.2$ guaranteeing the validity of the GP equation.  The intra-species scattering length can be reduced by a factor of two by the Feshbach resonance technique \cite{fesh} to obtain a quasi-2D nonlinearity of 5000. 
 Hence,
  with present experimental know-how,  these phase-separated vortex lattices can be generated and studied in
a laboratory using two hyperfine spin states of $^{87}$Rb.
 }

\section*{Acknowledgements}
\noindent

SKA thanks the Funda\c c\~ao de Amparo \`a Pesquisa do
Estado de S\~ao Paulo (Brazil) (Project: 
2016/01343-7) and the Conselho Nacional de Desenvolvimento Cient\'ifico e Tecnol\'ogico (Brazil) (Project:
303280/2014-0) for partial support.

%\section*{References}

% 
\end{document}